\documentclass[notitlepage,superscriptaddress, 11pt, nofootinbib]{revtex4-1}
\usepackage{graphicx}
\usepackage{preamble}
\usepackage{yquant,braket}
\usepackage{soul}
\usepackage{natbib}
\usepackage{booktabs}
\usepackage{makecell}
\usepackage{url,hyperref,lineno,microtype}
\usepackage{makecell}

\begin{document}

\title{The prospects of Monte Carlo antibody loop modelling on a fault-tolerant quantum computer} 

\author{Jonathan Allcock}
\email{jonallcock@tencent.com}
\affiliation{Tencent Quantum Laboratory}
\author{Anna Vangone}
\email{anna.vangone@roche.com}
\affiliation{Roche Pharmaceutical Research and Early Development, Large Molecule Research, Roche Innovation Center Munich, Penzberg, Germany}
\author{Agnes Meyder}
\affiliation{Roche Pharmaceutical Research and Early Development, Operations, Roche Innovation Center Basel, Basel, Switzerland}
\author{Stanislaw~Adaszewski}
\email{stanislaw.adaszewski@roche.com}
\affiliation{Roche Pharmaceutical Research and Early Development, Operations, Roche Innovation Center Basel, Basel, Switzerland}
\author{Martin Strahm}
\affiliation{Roche Pharmaceutical Research and Early Development, Operations, Roche Innovation Center Basel, Basel, Switzerland}
\author{Chang-Yu Hsieh}
\affiliation{Tencent Quantum Laboratory}
\author{Shengyu Zhang}
\email{shengyzhang@tencent.com}
\affiliation{Tencent Quantum Laboratory}

\begin{abstract}

Quantum computing for the biological sciences is an area of rapidly growing interest, but specific industrial applications remain elusive. Quantum Markov chain Monte Carlo has been proposed as a method for accelerating a broad class of computational problems, including problems of pharmaceutical interest.

Here we investigate the prospects of quantum advantage via this approach, by applying it to the problem of modelling antibody structure, a crucial task in drug development. To minimize the resources required while maintaining pharmaceutical-level accuracy, we propose a specific encoding of molecular dihedral angles into registers of qubits and a method for implementing, in quantum superposition, a Markov chain Monte Carlo update step based on a classical all-atom force field. We give the first detailed analysis of the resources required to solve a problem of industrial size and relevance and find that, though the time and space requirements of using a quantum computer in this way are considerable, continued technological improvements could bring the required resources within reach in the future.

\end{abstract}

\maketitle

\section{Introduction}

The last few years have seen a dramatic increase in global investment in quantum computing, accompanied by a concerted effort to find industrial applications that can outperform existing computational methods.  In the biological sciences, a number of initial studies on quantum computing for protein folding have been made. In particular, the works of~\cite{perdomo2008construction, perdomo2012finding, babbush2012construction,babej2018coarse, outeiral2021investigating} investigate simulating lattice-based models of proteins using analog quantum computers such as quantum annealers; For gate-based quantum computing, lattice-models of proteins have also been considered in~\cite{fingerhuth2018quantum,robert2021resource} using variations of the quantum approximate optimization algorithm (QAOA)~\cite{farhi2014quantum} and the variational quantum eigensolver (VQE)~\cite{peruzzo2014variational}. More realistic, non-lattice based models have also been studied, with~\cite{mulligan2020designing} using the D-Wave quantum annealer in conjunction with a Rosetta energy function and side-chain rotamer library, and \cite{casares2022qfold} proposing a method which combines the AlphaFold~\cite{senior2020improved} algorithm with quantum walks\footnote{Of the previous quantum studies on protein folding, the paper of~\cite{casares2022qfold} -- the preprint of which appeared while our manuscript was in preparation -- is closest to ours.  Like this manuscript, their method is also based on quantum Markov Chain Monte Carlo, but they do not consider the implementation of the key quantum operations, viewing these as being performed by a `black box'.}. While these works have made important contributions to our understanding of how quantum computers may be applied to this domain, no methods have yet been proposed that might solve specific problems to a speed and accuracy that would make them attractive to industry.

Motivated by the growing need to understand the true potential of quantum computing for solving real-world problems of industrial size and commercial relevance, here we show how classical Markov chain Monte Carlo (MCMC) methods based on torsion space conformation updates and all-atom force fields, such as those used in the Rosetta software package, can be adapted into a quantum computing procedure to predict the 3D structure of protein loops starting from their amino acid sequence.  As a potential application, we have in mind the modelling of antibody loops -- in particular, the H3 loop -- a crucial task in the development of therapeutic antibodies. This problem lies in the sweet spot of (i) being of practical importance to the pharmaceutical industry, as existing computational methods cannot predict H3 loop structures to the required near-atomic-level accuracy quickly enough to be part of an industrial workflow\footnote{While advances continue to be made in machine learning for protein folding -- most notably with the announcement of AlphaFold 2~\cite{deepmind2020alphafold2} in the CASP 14 protein structure prediction competition -- not all protein folding problems can be solved with these new methods, and there is the need to continue to explore the potential of quantum computing for this domain.}; and (ii) involving molecules of a size (typically $3$ to $30$ amino acid residues long) that, as we show, the problem can be tackled on a quantum computer with resources that are plausibly within reach in the future. 

 Structurally, antibodies consist of two identical pairs of polypeptides chains,  with each pair comprising a heavy chain (containing approx. 500 amino acid residues) and a light chain (approx. 200 amino acids) Their ability to bind to a large variety of molecular targets with high affinity and specificity has led to antibodies becoming the predominant class of new therapeutics and diagnostics tools in recent years. The function of antibodies, together with their desired drug profile (e.g. affinity, stability, half-life, tissue penetration \cite{kim2005antibody}), is a direct consequence of their structure. As experimental structure determination of antibodies is time-consuming and costly, computational structure prediction plays a crucial role in accelerating and facilitating the development of antibody-therapeutics.

\begin{figure}
    \centering
    \includegraphics[height=5cm]{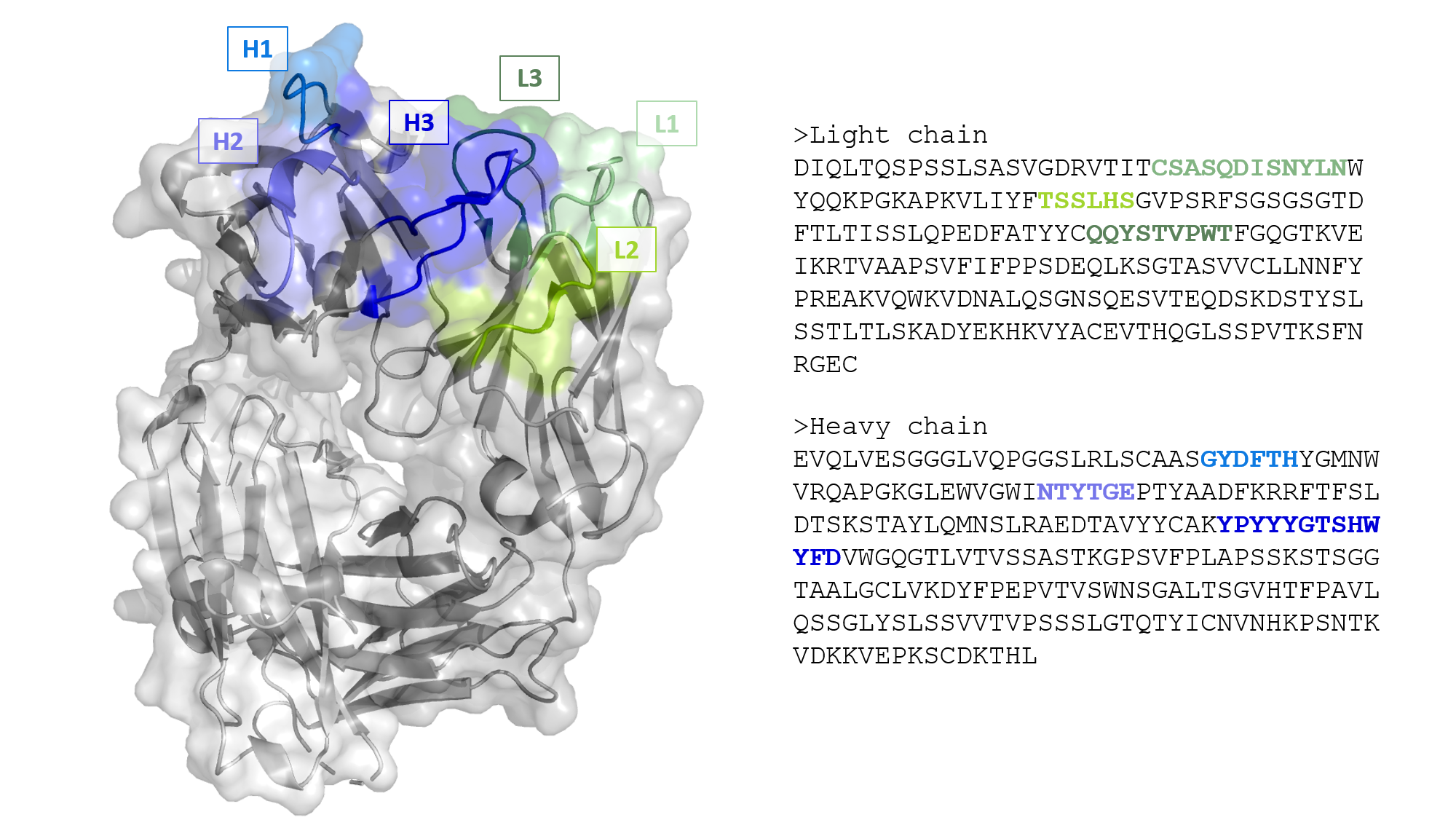}
    \caption{(Left) Cartoon and surface representation of the Fab region of the anti-VEGF antibody with PDB code 1CZ8. Framework is represented in gray, complementarity-determining regions in green (light chain loops L1, L2 and L3) and blue (heavy chains loops, H1, H2 and H3). (Right) The corresponding amino acid sequences of the chains. Image created with PyMOL \cite{pymol}.}
    \label{fig:cdrs}
\end{figure}

Antibody binding occurs via a specific antigen-binding region, characterized by 6 hypervariable loops -- called complementarity-determining regions (CDRs) -- located on the variable domains of the light (L1, L2, L3) and heavy (H1, H2, H3) chains (Fig.\ref{fig:cdrs}). While antibodies are typically more rigid and stable than other proteins, they are known to retain a certain amount of plasticity to accommodate for different antigens, with a degree of flexibility inherent to many CDRs~\cite{fernandez2019cdr, fernandez2020antibodies, fernandez2021ensembles}. For 5 of the CDRs loops (L1, L2, L3, H1 and H2) though, limited shapes have been observed, leading to the identification of definite canonical structures based on their sequences. In contrast, H3 loops exhibit high diversity -- with longer loops typically displaying more conformal variety -- and cannot be classified into canonical groups. Consequently, the main antibody modelling problem of interest is the accurate prediction of the H3 loop. While atomic resolutions (i.e. $1$ Angstrom $=10^{-10}$m) can be reached for canonical CDRs, accuracy ranging from $1.5$-$3$ Angstroms or worse can be expected for H3.  Furthermore, such loops do not exist in a unique structural conformation, but rather as an ensemble of different states that can occur on different timescales and with different probabilities. As experimental structures for antibodies are, in most cases, derived by X-ray crystallography at low temperature ($\sim100$K) at which only the most dominant conformation can be observed, other metastable states of the loops which are present at physiological temperatures must often be deduced via simulation. Thus, it is desirable and insightful for H3 loop modelling procedures to output not only a single candidate H3 structure, but rather sample from the Boltzmann distribution of possible structures.

Classical MCMC methods for loop modelling sample from the Boltzmann distribution based on a chosen state space $\Omega$ of possible configurations of the molecule. Starting from an initial configuration $x\in\Omega$ of the molecule, the following steps are then iterated:
\begin{enumerate}
    \item \quad Propose a random update $x\rightarrow x'\in\Omega$.
    \item \quad Accept the update with probability $\min\left\{1,\frac{e^{-E(x')/ T}}{e^{-E(x)/ T}}\right\}$.
\end{enumerate}
Here, $E(x)$ is the energy of a configuration and $T$ is the temperature. This procedure -- if run for a sufficiently long time -- is guaranteed to converge to a configuration drawn from the distribution $p(x)\sim e^{-E(x)/T}$, and has effectiveness that depends on defining a state space that can capture the biologically relevant structures at appropriate levels of granularity, and update rules that can efficiently explore this space, avoiding spending too long stuck in energetically unfavorable configurations.  In practice, the accurate MCMC modelling of pharmaceutically relevant loops can take days to weeks to complete -- too long to be part of a commercially feasible workflow.

Here we are interested in using quantum computing to accelerate the MCMC process, and give the first detailed analysis of the resources required to solve a protein structure problem of industrial size and relevance. 
Our focus is specifically on fault-tolerant gate-based quantum computing (see Sec.~\ref{sec:circuit-model}). While it may be several decades before large-scale fault-tolerant devices are available ~\cite{sevilla2020forecasting}, they are widely believed to offer the best long term prospects for practical advantage of classically-intractable problems, with mathematically provable efficiency of algorithms possible in some cases.  In contrast, other models of quantum computing may be available sooner, but may not guarantee the same long-term advantages. Analog approaches to quantum computing such as adiabatic quantum computing~\cite{farhi2000quantum}, quantum annealing or continuous time quantum walks~\cite{farhi1998quantum} lack practical means of error-correction, which may limit the size of computations that can be performed; and the embedding of computational problems in a form amenable to annealing makes the application of realistic energy models challenging (e.g.,~\cite{marchand2019variable, mulligan2020designing} require non-trivial procedures interleaving classical and quantum computation). Hybrid quantum-classical approaches such as VQE and QAOA, which are gate-based but primarily targeted at noisy, non-error-corrected quantum computers, similarly lack strong evidence for practical advantage or scalability. For an overview on the prospects of various quantum computing technologies see~\cite{national2019quantum}.

\section{Methods}

\subsection{Ciruit model of quantum computation}\label{sec:circuit-model}
In the standard circuit model of quantum computing, computational tasks are carried out by applying operations, known as gates, to registers (i.e., groups) of qubits.  At the end of a sequence of gates, one or more of the qubits are measured and the results recorded. The size and complexity of the quantum circuit required to solve a particular task determines the overall algorithmic running time and, in particular, whether or not an advantage can be gained over existing classical computational methods.  As individual qubits and gates are invariably error prone, quantum error correction procedures must be applied for long circuits to be computed. The aim of error correction is to use multiple noisy physical qubits to encode a single error-free \emph{logical} qubit, which comes at the cost of additional qubits and computational time. As the error correction operations themselves may be faulty, one must take care to ensure that the net effect is an overall reduction in error. This is referred to as fault-tolerance and, if achieved, can be used to drive errors arbitrarily low, enabling large scale computations to be implemented.  For further background, we refer readers to~\cite{outeiral2021prospects} for a good introduction to quantum computing from a biological sciences perspective. In the appendix we give additional details on topics specific to this work,  including accelerating MCMC via quantum walks, resource overheads required for error correction, and evaluating complicated functions in a quantum circuit.

\subsection{Quantum MCMC}

Quantum Markov chain Monte Carlo~\cite{szegedy2004quantum} is an approach to speeding up classical MCMC methods on a fault-tolerant quantum computer (FTQC).  By encoding the state of the system of interest in a number of qubits and translating the update and acceptance rules into a sequence of quantum gates, the number of update steps required can be reduced to roughly the square root of the number of steps required classically (see Appendix~\ref{app:quantum-mcmc}).  While this quadratic reduction provides an opportunity for quantum advantage, the time required for each step may be longer in the quantum case, and care is needed in analyzing whether a speedup can be obtained. Furthermore, the success of the approach depends, among other things, on finding an efficient quantum encoding of the 3D structure of the antibody loop. That is, a way of representing the antibody structure in the state of multiple qubits. 

To evaluate the feasibility of quantum MCMC for antibody loop modelling, we propose a specific encoding of molecular dihedral angles into registers of qubits and a method for implementing the MCMC update step coherently in quantum superposition. To enable the latter, we propose a quantum subroutine (Quantum SN-NeRF) based on the classical Self-Normalizing Natural extension Reference Frame~\cite{parsons2005practical} method for coherently converting from  dihedral angles to Cartesian coordinates. We estimate the number of qubits and time required to implement such an approach on an FTQC and find that, while there are limited prospects for an advantage on a first generation FTQC, continued technological improvements could bring the required resources within reach on future quantum devices.

\subsection{Dihedral angles}

We consider polypeptides consisting of $L$ amino acid residues, containing $N$ heavy (non-hydrogen) atoms. Atomic positions can be described by Cartesian coordinates in 3D space, or relative to one another using dihedral (or torsion) angles.  While the Cartesian representation is convenient for computing atomic forces and determining deviations of predicted atomic positions from experimentally determined positions, the dihedral representation can be preferable for generating perturbations to the molecular structure.

\begin{figure}[h]
    \centering
    \includegraphics[width=0.6\textwidth]{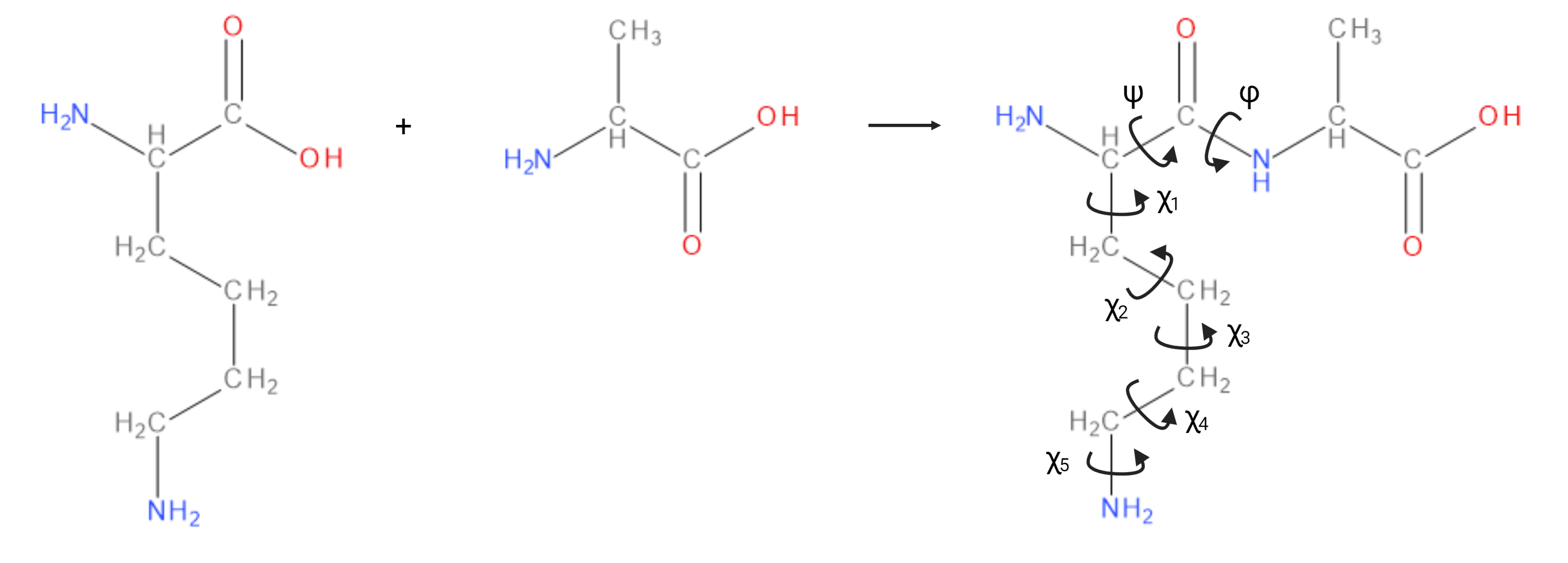}
    \caption{The amino acids leucine and alanine forming a dipeptide. The backbone angles $\varphi$, $\psi$, and side chain angles $\chi_i$ of alanine are annotated. In any polypeptide, $\varphi_i$ is defined by atoms \ch{C_{i-1}} - \ch{N_i} - \ch{C$\alpha_{i}$} - \ch{C_i}. $\psi_i$ is defined by atoms \ch{N_i} - \ch{C$\alpha_{i}$} - \ch{C_i} - \ch{N_{i+1}}, and $\omega_i$ is defined by atoms  \ch{C$\alpha_{i-1}$} - \ch{C_{i-1}} - \ch{N_i} - \ch{C$\alpha_{i}$}. The  $\omega_i$ angle is not shown in the figure as it is nearly always close to $180^\circ$.} Image created with BIOVIA Draw and Biorender.com.
    \label{fig:amino-acid}
\end{figure}

In the dihedral formalism, any four consecutive backbone atoms \ch{A}-\ch{B}-\ch{C}-\ch{D} define two planes, containing \ch{A}-\ch{B}-\ch{C}, and \ch{B}-\ch{C}-\ch{D} respectively. The angle between these two planes is the associated dihedral angle and, in a polypeptide backbone, each residue has three associated dihedral angles labelled $\varphi$, $\psi$ and $\omega$. A complete internal representation of the backbone is given by specifying each of the dihedral angles and bond lengths between consecutive backbone atoms, as well as the bond angles between any three consecutive atoms. The side chains of a polypeptide can similarly be described by dihedral angles $\chi_i$, where $i=1,2,3,\ldots$ depending on the length of the side chain (see Fig.~\ref{fig:amino-acid}).

\subsection{Design considerations}

\begin{figure}
    \centering
    \includegraphics[height=8cm]{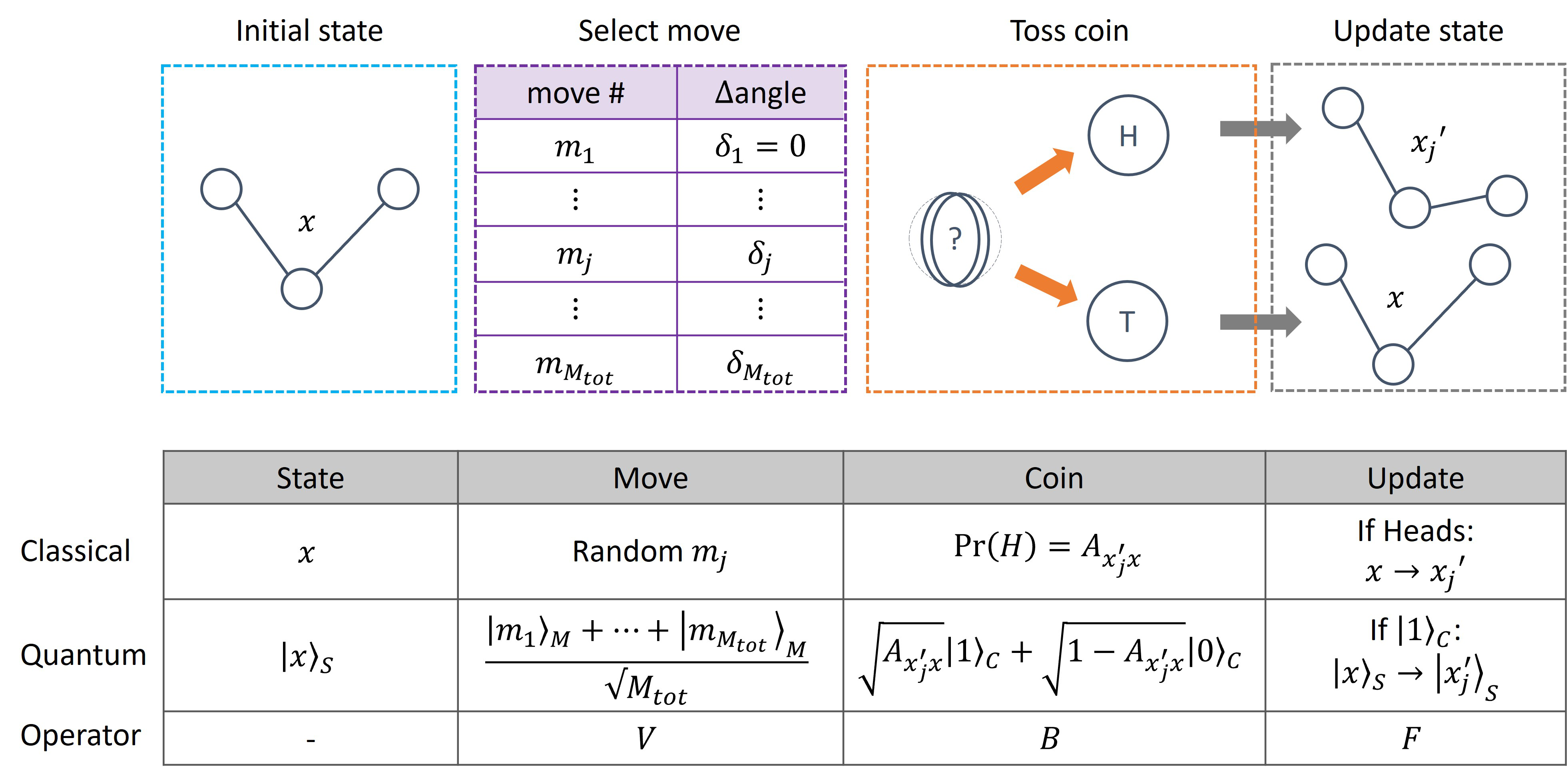}
    \caption{(Top) Schematic of the (classical and quantum) MCMC process, illustrated with a toy model consisting of a 3 atom molecule defined by a single angle $x$. Given an initial state of the system, an update move is proposed from a list of possible moves.  In this example, a move $m_j$ corresponds to increasing the angle $x$ by an amount $\delta_j$. A coin is then tossed and, if a heads is obtained, the state is updated to $x_j'$.  (Bottom) Classical vs. quantum implementation of this process, and the quantum registers ($S,M,C$) and operators ($V,B,F$) involved in implementing each step. In addition, there is a fourth quantum operator ($R$), which has no classical analogue, but is required for the quantum algorithm to obtain an asymptotic speedup.  }
    \label{fig:qwalk-schematic}
\end{figure}

Given a classical Markov chain with state space $\Omega$, let $\+M =\{m_1, m_2, \ldots \}$ be a set of moves, i.e. mappings $m_j :\Omega \ra \Omega$ that correspond to proposed state space updates $x\ra x_j' = m_j(x)$, and let $P_{M}$ be a probability distribution over the moves. That is, when the chain is in state $x$, the update $x\ra x'_j$ is proposed with probability $T_{x'_jx} = P_M(j)$. Once proposed, the update is accepted with probability given by the Metropolis-Hastings (MH) rule:
\begin{linenomath}
\eql{
A_{x' x} = \min\left\{1,\frac{e^{-E(x')/ T}}{e^{-E(x)/ T}}\right\} \label{eq:MH}
}
\end{linenomath}
In our approach, we follow the LHPST~\cite{lemieux2020efficient} quantum MCMC framework, where a quantum walk operator $W$ is implemented by four other quantum operators $(V,B,F,R)$, acting on System (S), Move (M) and Coin (C) quantum registers. Mathematical definitions of these operators are given in Appendix~\ref{app:loop-resources}, but their functions, and that of the various registers, can be understood by analogy with the classical MCMC process.  In what follows we use Dirac notation $\ket{\cdot}$ to denote a quantum state, with a subscript, e.g., $\ket{0}_C$, denoting the register. Both classical and quantum MCMC are based on the same principles: given an initial state, (i) propose an update move; (ii) toss a biased coin; (iii) if a heads is obtained, update the state based on the proposed update.  The difference is in how these steps are implemented (see Fig.~\ref{fig:qwalk-schematic}). In the classical case, the update move is chosen randomly from a list of possible moves, and the probability of obtaining a heads is given by the MH rule. In the quantum case, the state of the system is stored in the $S$ register. Then, rather than selecting a single random update rule, the $V$ operator is used to create a superposition of all possible update moves in the $M$ register. The effect of the coin toss is implemented by the $B$ operator, which creates a superposition of $\ket{0}_C$ and $\ket{1}_C$ in the coin register, where the probability of measuring $\ket{1}_C$ is given by the MH rule. The $F$ operator acts to update the state if the coin register is in the $\ket{1}_C$ state (i.e., Heads).  In addition, a fourth quantum operator (the $R$ operator) is needed, which serves to reflect the move and coin states if both of them are zero, i.e. $\ket{0}_M\ket{0}_C\ra -\ket{0}_M\ket{0}_C$. This has no classical analogue, but is required for the quantum algorithm to provide an asymptotic speedup.

 The Coin register consists of a single qubit, while the number of qubits in the System and Move registers is problem-dependent. The quantum walk operator $W$ is then defined as
\begin{linenomath}
\eql{
W &= RV^\dagger B^\dagger  F B V \label{eq:W-tilde}
}
\end{linenomath}

To apply this framework to the problem of antibody loop modelling, we make the following design choices:

\noindent\textbf{Dihedral angle encoding.} We represent each amino acid residue in the loop by its backbone and side chain dihedral angles, assuming ideal bond lengths and angles. Lookup tables of these angles can be constructed by sampling data from Ramachandran plots~\cite{ramachandran1963stereochemistry} and backbone-dependent rotamer libraries, e.g.~\cite{shapovalov2011smoothed}. By specifying an index into the tables for each residue, the biologically relevant structures of the loop can be given in compact form for encoding in a register of qubits. 

\noindent\textbf{Update rules.} The dihedral angle encoding allows for an efficiently implementable Monte Carlo update step corresponding to replacing a randomly chosen backbone or side chain dihedral angle with another randomly chosen value from the corresponding lookup table\footnote{The lookup table approach to encoding and updating can be extended to other schemes based on dihedral angles, e.g. the backbone fragment insertion scheme of Rosetta where a multiple-residue fragment is randomly selected and the associated torsion angles replaced with the torsion angles from another fragment from a precomputed list.}.

\noindent\textbf{Energy function.} We are interested in high-accuracy modelling and thus take $E(x)$ to be specified by a classical all-atom force field such as CHARMM36m~\cite{huang2017charmm36m}.

\noindent\textbf{Conversion to 3D coordinates.} Evaluation of $E(x)$ necessitates conversion of the dihedral angle representation of the loop into 3D coordinates. To do so, we first use a Quantum Read-Only Memory (QROM)~\cite{babbush2018encoding} approach to convert the dihedral lookup table indices into their corresponding angles. Then, we adapt a variant~\cite{parsons2005practical} of the classical Natural-extension Reference Frame (NerF) algorithm used by the Rosetta software package to give a quantum procedure (QSN-NeRF) for  coherently converting from dihedral angles to 3D coordinates in quantum superposition.

\noindent\textbf{Energy calculation and Metropolis update.} After conversion to 3D coordinates, the energy of the existing and proposed configurations can be computed. Implementing the Metropolis update in the LHPST framework requires evaluating $\arcsin \lp {\sqrt{\min\{1, e^{- (E(x')-E(x))/T}\}}}\rp$ in quantum superposition.  Due to the high costs of implementing certain quantum mathematical operations, we divide this calculation into two parts; (i) We compute $E(x), E(x')$ by decomposing the energy function into elementary arithmetic operations for which efficient quantum circuits are known; (ii) following~\cite{sanders2020compilation} we propose implementing the subsequent $\arcsin\lp \sqrt{\cdot}\rp$ computation via  QROM lookup.

With these design choices we obtain estimates of the resources required to implement the LPHST quantum walk given in Table~\ref{tab:resources-summary} (see Appendix~\ref{app:loop-resources} for details).  In leading candidate proposals for fault-tolerant quantum computing such as via the surface code, the Toffoli (controlled-controlled-NOT) gate is expected to take orders of magnitude longer to implement than other gates, as each Toffoli gate first requires the production of an associated \emph{magic} state via an expensive process known as distillation. We therefore estimate the computational time and number of qubits required to implement the various quantum operators required by our approach by the number of Toffoli gates needed.

\begin{table}
\centering
\resizebox{0.9\textwidth}{!}{%
\begin{tabular}{l|c|c|c}
	 & logical qubits & temporary qubits & Toffoli count\\
	\hline
	System Register & $2  L b_T$ & & {} \\
	\hline 
	Move Register & $b_T + \log{L} + 1$ & &{} \\
	\hline 
	Coin Register & $1$ & & \\ \hline
	Operator $V$ & {} & 0 & 0 \\
	\hline 
	Operator c$F$ & & $\log L + 1$ &  $8L - 4 + 2Lb_T$ \\
	\hline 
	Operator c$R$ &  &$\log L + b_T + 1$ &  $2\left(\log{L}+b_T+1\right)$ \\
	\hline 
	Operator $B$ & &{} &  \\ 
	
	\quad Quantum SN-NeRF & & $96N b$ &  $N (147 b^2 + 34 b)$ \\
	
	\quad \makecell[l]{Force field non-bonded} & &$20N(N-1)b$  &  $N(N-1)(31b^2 + 8b)$ 
\end{tabular}%
}
\caption{Resources Summary. $L$=number of residues, $N$=total number of heavy atoms in loop, $b_T$=number of bits in lookup table keys, $b$=number of bits of precision used to store Cartesian coordinates. $B$ operator resources are estimated from contributions from quantum SN-NeRF and non-bonded energy terms. Values for quantum SN-NeRF and non-bonded force field terms are double the numbers given in Appendix~\ref{app:loop-resources}, as the computations are performed for the current and proposed update states.}
\label{tab:resources-summary}
\end{table}
\label{sec:statespace}

\section{Results}

To benchmark our approach and understand the technological requirements necessary for a quantum advantage, we consider the MKHMAGAAAAGAVV H1 loop from the Syrian hamster prion protein, which was modelled to high accuracy via classical Monte Carlo methods in~\cite{ulmschneider2006monte} using the OPLS-AA all-atom force field, GBSA implicit solvent model and concerted rotation updates. While some of these details differ from our quantum procedure, and the loop considered is an H1 rather than H3 loop, the loop size ($L=14, N=88$) and the number of classical MCMC steps ($10^{6}$ per structural sample) serve as a useful baseline for comparison.  Estimates of the resources required to solve the same problem using our quantum approach are given in Table~\ref{tab:running-time} (see Appendix~\ref{app:benchmarking} for more details). The number of qubits needed and the quantum running time per step are based on a superconducting quantum processor running the surface code~\cite{bravyi1998quantum, dennis2002topological}, a leading candidate for error-corrected quantum computing.  As mentioned, for such a system, the resource bottleneck is the time and qubits required for the magic state distillation used to implement non-Clifford operations such as the Toffoli gate.  

Using the state-of-the-art $\ket{CCZ}$ distillation factory of~\cite{gidney2019efficient}, and assuming first generation large-scale FTQC have physical error rates of $10^{-4}$~(see~\cite{sevilla2020forecasting} for forecast timelines of quantum computing technology) and a surface code cycle time of $1\mu s$\footnote{Cycle times of this order of magnitude have essentially already been demonstrated experimentally~\cite{chen2021exponential,zhao2021realizing}.}, a single Toffoli gate can be distilled every $170.5 \mu$s using a specially reserved section of the quantum computer -- known as a factory -- consisting of $\sim 1.3\times 10^5$ physical qubits. In the long run, surface code cycle times of $200$ns should be achievable~\cite{fowler2012surface}.  If physical error rates can be reduced to $10^{-5}$ then the required resources drop considerably, with a single Toffoli distillable in $12.1 \mu$s using a factory consisting of $\sim 2\times 10^4$ physical qubits, and the number of physical qubits required to encode a single logical data qubit reducing from roughly $1,000$ to $500$ (see Appendix~\ref{app:Toffoli}).

\begin{table}
\centering
\resizebox{\textwidth}{!}{%
\begin{tabular}{c|ccc|ccc|c}
    & TF & $p_e$ & $t_S$& $n_{\text{steps}}$/sample& $t_{\text{sample}}$ & time / $10^3$ samples & physical qubits \\ \hline
    Classical & - & -& -&$10^6$ & $230$ s & $2.7$ \text{ days}  & - \\ \hline
    Gen.1 & $1$ & $10^{-4}$ & $1\mu$s&$1.4\times 10^3$& $7.5$ years & $7,500$ years & $4\times 10^6$  \\
    Gen.1  & $100$ & $10^{-4}$& $1\mu$s&  $1.4\times 10^3$& $20.6$ days & $57$ years & $2\times 10^7$  \\ \hline
    Future  & $100$ & $10^{-5}$& $200$ns & $1.4\times 10^3$  & $16.5$ hours  & $687$ days &  $10^8$ \\
    Future & $1000$ & $10^{-5}$& $200$ns & $1.4\times 10^3$ &  $1.3$ hours &  $55$ days &  $10^9$
\end{tabular}%
}
\caption{Resource estimates for the $14$ residue MKHMAGAAAAGAVV antibody loop, giving the number $n_{steps}$ of MCMC steps required per structural sample, the time $t_\text{sample}$ required per sample, and the total time required to obtain $10^3$ samples. The first row corresponds to the classical MCMC result from~\cite{ulmschneider2006monte}. Quantum estimates correspond to different assumptions on physical error rates $p_e$ and surface code cycle times $t_S$. For first generation (Gen.1) large scale FTQC, we assume $p_e = 10^{-4}, t_S = 1\mu$s. For future quantum devices, $p_e = 10^{-5}, t_S = 200$ns are plausibly achievable \cite{fowler2012surface,sevilla2020forecasting}.  The TF column indicates the number of parallel Toffoli distillation factories that are assumed to be available.}
\label{tab:running-time}
\end{table}

In terms of qubit numbers, we assume that the first generations of large-scale fault-tolerant devices will be limited to $O(10^6)$ to $O(10^7)$ physical qubits, but that these numbers may increase by one or two orders of magnitude in the longer term. Table~\ref{tab:running-time} gives estimates based on these first generation and future FTQC computing parameter regimes. As the number of Toffoli gates required is significant, a single distillation factory may not suffice, and we thus give estimates assuming parallel access to varying numbers of factories.  The single step time estimates are based on the Toffoli gate count required to implement the quantum SN-NeRF conversion from dihedral to 3D coordinates, and the quantum circuit computation of the non-bonded energy terms in the force field (assuming no cut-off radius).  These steps dominate the classical computation time and, as the form of the non-bonded terms is similar to those in other popular force fields, do not tie our results to a specific force field. The number of qubits includes additional ancillary registers required to store temporary arithmetic values prior to uncomputation, but ignores ancilla used in various quantum arithmetic primitives which are highly implementation-specific and depend on specific choices of quantum arithmetic circuits. These simplifications are sufficient for our goal of understanding the order of magnitude of technological performance required to obtain a quantum advantage. We find that, in spite of the careful design decisions made to minimize the resources required, first generation FTQC are unlikely to provide an advantage over classical MCMC techniques, with estimated computational times that greatly exceed those required classically.  However, with plausible improvements to physical qubit error rates and error correction speed, future quantum devices, with sufficient qubit numbers, may close the gap with existing MCMC approaches to the point where quantum walk methods may be competitive.  Further improvements to both hardware and algorithm design could eventually yield an overall quantum advantage. 

\section{Discussion}

In this work, we set out to understand the feasibility of quantum computing to accelerate MCMC for antibody loop modelling, and proposed a suitable state space and update rule for such a computation.  Our method is based on a dihedral angle encoding of the atoms involved, and a procedure (Quantum SN-NeRF) for coherently converting from dihedrals to Cartesian coordinates so that the classical force field potential energy function of each configuration can be evaluated in quantum superposition. 

While the encoding and conversions can be carried out efficiently, the energy function evaluation is costly and imposes a trade-off between the quadratic reduction in number of MCMC steps needed for the quantum approach, and a significant constant factor increases in the time required for each step. These long step times are in large part due to the time needed to implement fault-tolerant Toffoli gates used to carry out basic arithmetic operations.  We find that for system sizes of practical interest, this trade-off is not yet in favor of quantum computers, and that further developments in both algorithms and quantum hardware are likely needed in order for quantum computing to be practical in this domain. 

Our analysis indicates limitations of directly applying the quantum walk approach to classical MCMC methods and suggests that, without several orders of magnitude increases in fault-tolerant hardware efficiency, new quantum algorithms or design improvements to our scheme (e.g., alternative encodings of molecular states in qubits) will be needed to make protein folding practical on quantum computers.   These findings are in line with those in \cite{babbush2020focus}, which show significant challenges for constructing efficient quantum solutions to a number of non-toy-model optimization problems.  An interesting open problem is to investigate whether new force fields can be designed (or indeed, machine-learned~\cite{botu2017machine, unke2021machine}) to be efficiently implementable on quantum computers while still delivering sufficiently accurate results.

As investment and hype continue to grow in quantum computing, for meaningful and informed progress to be made, detailed analyses of specific problems facing industry must continue to be carried out and disseminated even (and indeed especially) if they show limitations or challenges with quantum computing. It is our hope that the results presented here shed some light on the future prospects of quantum computing as a tool for modelling antibody loops, and serve as a useful starting point for further improvements.  As a first step in this direction, our approach can be further refined, for instance by accounting for the presence of solvents, or constraining the ends of the loop at fixed anchor points (e.g. by kinematic closure). In addition, our proposed method can be applied to more general protein folding problems beyond antibody loop modelling, which may have different time and accuracy requirements for demonstrating quantum advantage. As quantum technology continues to improve, its viability as a competitive resource for the pharmaceutical industry will need to be continually reassessed.

\section*{Acknowledgments}
We would like to express our gratitude to Guy Georges, Alexander Bujotzek, Hubert Kettenberger, Detlef Wolf, Yvonna Li, Xavier Lucas, Bryn Roberts and Mari\"{e}lle van de Pol for their expertise, support, and encouragement throughout the course of this project. JA is grateful to Yicong Zheng for many helpful discussions and feedback during the preparation of this manuscript.

\bibliographystyle{unsrtnat} 
\bibliography{protein-bib.bib}

\appendix

\section{Accelerating MCMC via quantum walks} \label{app:quantum-mcmc}
Given a classical Markov chain with transition matrix $P$ with eigenvalues $1 = \lambda_1 > \lambda_2 > \ldots \ge \lambda_n > -1$, spectral gap $\Delta\defeq \lambda_1 - \lambda_2$, and stationary distribution $\pi$, one can define an associated quantum walk operator $W$ with several nice properties~\cite{szegedy2004quantum} . First, the eigenvalues of $W$ are $e^{\pm i \arccos(\lambda_j)}$, which satisfies $\abs{\arccos(\lambda_1) - \arccos(\lambda_2)}\approx \sqrt{2\Delta}$.  Second, the quantum state $\ket{\pi}\ket{0} = \sum_x \sqrt{\pi_x}\ket{x}\ket{0}$ is the unique $\lambda_1 = 1$ eigenstate of $W$.
Since measuring $\ket{\pi}$ in the computational basis gives a sample drawn from $\pi$, the following is a procedure for generating samples from the stationary distribution: (i) Create an initial quantum state $\ket{\psi_0}$ which has non-negligible overlap with $\ket{\pi}$, i.e. $\abs{\langle \pi\vert \psi_0\rangle} = q$. (ii) Perform quantum phase estimation of $W$ on $\ket{\psi_0}$. If the measured approximate phase $\tilde{\lambda}$ is closer to $\lambda_1=1$ than $\lambda_2 = 1-\Delta$ then you have successfully created $\ket{\pi}$, which can then be measured to give the desired sample.  This will happen with probability at least $q^2$ provided that $\tilde{\lambda}$ can be resolved to accuracy $\sqrt{2\Delta}/2$, which can be achieved in time  $\sqrt{\frac{2}{\Delta}}T(W)$, where $T(W)$ is the time required to implement a controlled-$W$ operator. This is quadratically faster than classical MCMC methods, which require time proportional to $\frac{1}{\Delta}$ to mix sufficiently that a good sample is obtained \cite{levin2017markov}.  A quantum advantage can thus be obtained provided that $T(W)$ is not too large.


\section{Error rates, Toffoli count, distillation and logical qubits}\label{app:Toffoli}

While the quality of quantum operations is commonly measured by a metric known as the gate fidelity~\cite{nielsen2002simple}, for large-scale error-corrected quantum computation, a more relevant metric is the error rate of single and two-qubit gates. Qualitatively, this is a measure of the probability that the output of a quantum gate differs from its ideal, intended behaviour. The relationship between fidelity and error rate can be complicated and counter-intuitive, with high fidelities not-necessarily translating to low error rates~\cite{sanders2015bounding}. The importance of the error rate comes from the threshold theorem~\cite{shor1995scheme, aharonov2008fault,knill1998resilient, knill2005quantum} of quantum computing, which guarantees that if the error rate is below a certain value then arbitrarily long computations can be performed with only minimal resource overheads. Current state-of-the-art superconducting processors are able to achieve two-qubit error rates (which are typically much larger than single qubit error rates) of order $10^{-3}$, and rates of $10^{-4}$ and $10^{-5}$ have been optimistically forecast as achievable by the early 2030s and 2040s, respectively~\cite{sevilla2020forecasting}.

Using the state of the art $\ket{CCZ}$ distillation factory of~\cite{gidney2019efficient}, a single Toffoli gate can be distilled in $5.5d$ surface code cycles, and $2(12d\times 6d)$ physical qubits, where $d$ is the code distance. For physical error rates of $10^{-4}$ a code distance $d=31$ is sufficient to perform quantum circuits of length required by our quantum MC algorithm\footnote{The resource estimation spreadsheet included in the supplementary information of~\cite{gidney2019efficient} was used to derive the estimates presented in this section.}. A surface code cycle time of $1 \mu$s (times of this order of magnitude have already been demonstrated experimentally~\cite{chen2021exponential,zhao2021realizing}) gives a distillation rate of one Toffoli gate every $170.5 \mu$s, using a factory consisting of approximately $130,000$ physical qubits. In the long run, surface code cycle times of $200$ns could, in principle, be achievable~\cite{fowler2012surface}.  If physical error rates can be reduced to $10^{-5}$ then the required code distance drops to $d=11$, and a single Toffoli can be distilled in $12.1 \mu$s using a factory consisting of approximately $20,000$ qubits.

In the surface code on a superconducting system, the number of noisy, error-prone physical qubits required to encode a single, fault-tolerant, logical qubit scales as $2d^2 (1 + R)$, where the factor of $2$ accounts for measurement qubits, and $R$ is an overhead parameter used to allow for space to route braiding operations in the surface and which we take to be $0.5$.  Note the distance $d$ used in encoding the logical data qubits can differ from the distance parameter used in the distillation process. For first generation and future FTQC we take code distances $d=19$ and $d=13$, respectively, corresponding to $1,083$ and $507$ physical qubits per logical qubit. 

The number of Toffoli gates (to leading order) required to implement common arithmetic operations is given in Table~\ref{tab:toffoli-count}. Multiple proposals for quantum arithmetic circuits exist~\cite{vedral1996quantum, draper2000addition, cuccaro2004new, draper2004logarithmic, takahashi2009quantum, bhaskar2015quantum, munoz2018t, gidney2018halving, haner2018optimizing, cao2013quantum, wang2020quantum, sanders2020compilation}, each with slightly different resource requirements. For our purposes, the leading-order ballpark figures given in the Table suffice. Some of these schemes require ancillary registers as a temporary working space. To simplify our analysis we ignore these ancillary requirements.

\begin{table}
\centering
\begin{tabular}{c|c|c}
 Operation & \text{Code} &$\+T$ \\ \hline
 $x + y$ & \text{ADD} & $b$ \cite{gidney2018halving} \\
  $x^2$ & \text{SQR} & $b^2 /2$ \cite{sanders2020compilation}\\
 $x \times y $ & \text{MUL}& $2b^2$ \cite{sanders2020compilation} \\
 $1/\sqrt{x}$ & \text{INVSQRT}&$18 b^2$ \cite{haner2018optimizing} \\
\end{tabular}
\caption{Number of Toffoli gates $\+T$ (to leading order) required to implement common quantum arithmetic operations on $b$-bit operands. The value for $\frac{1}{\sqrt{x}}$ assumes $2$ Newton iterations in the scheme of \cite{haner2018optimizing}.  }
\label{tab:toffoli-count}
\end{table}

\section{Quantum function evaluation}

Computing a function $f(x)$ in quantum superposition and storing the results in register, i.e. effecting the transformation
\begin{linenomath}
\eql{
\sum_j a_j \ket{x_j}\ket{0}&\rightarrow \sum_j a_j \ket{x_j}\ket{f(x_j)} \label{eq:qfe}
}
\end{linenomath}
can be carried out as follows: (i) Decompose $f$ into a computational graph of elementary arithmetical steps. (ii) Implement quantum circuits for carrying out each of these elementary steps reversibly. (iii) Each additional step requires new qubits to be added to the system. To control the total number of qubits required, previously used qubits are freed by periodically uncomputing -- performing inverse quantum circuits -- certain steps along the way.  While this prevents the number of qubits required from continuously growing, the time required for the uncomputation lengthens the total time required for evaluating $f$. For a computation consisting of $\ell^n$ steps (each taking unit time) with results stored in temporary registers, uncomputation of the registers every $\ell$ steps requires a total of $n(\ell-1)+1$ ancillary registers, and increases the computational time by a power of $\log(2\ell-1) / \log \ell$~\cite{bennett1989time}.

Since implementing function evaluation in a quantum circuit can be costly, when high accuracy results are not required, less resource-intensive approximate methods may suffice.  In particular, for simple functions of only one or two input variables, an approach based on Quantum Read-Only Memory (QROM)~\cite{babbush2018encoding} can be used. In this method, values of a function $f$ are pre-computed classically at discrete values $j$, and the $(j, f(j))$ pairs are stored in a lookup table. A quantum circuit can then be compiled which allows for the coherent access of values in the table, i.e. it enables the transformation
\begin{linenomath}
\eql{
\sum_{j} a_j\ket{j}\ket{0}&\rightarrow \sum_{j} a_j\ket{j }\ket{f(j)} \label{eq:qrom},
}
\end{linenomath}
where a table of length $k$ can be implemented with a Toffoli gate count of $k-1$, and independent of the number of bits used to represent each value $f(j)$ of the database. Eq.\eqref{eq:qrom} is equivalent to function evaluation restricted to points corresponding to the indices $j$.  If one wishes to evaluate the function at more general points, then a linear interpolation scheme can give an approximate solution with some additional overhead~\cite{sanders2020compilation}.

\section{Quantum resources for loop modelling}\label{app:loop-resources}

\subsubsection{LHPST quantum MCMC framework}

The LHPST~\cite{lemieux2020efficient} quantum MCMC framework is based on a quantum walk operator $W$ implemented by four other quantum operators, acting on System (S), Move (M) and Coin (C) quantum registers. These are defined mathematically as
\begin{linenomath}
\eql{
V&:\ket{0}_M \rightarrow \sum_{j=1}^{\abs{\+M}} \sqrt{P_M(j)}\ket{j}_M \\
B&:\ket{x}_S\ket{j}_M\ket{0}_C \rightarrow \ket{x}_S\ket{j}_M \lp \sqrt{1-A_{x'_j x}}\ket{0}_C+ \sqrt{A_{x_j'x}}\ket{1}_C\rp \label{eq:B-op}\\
F&:\ket{x}_S\ket{j}_M\ket{c}_C \rightarrow  \begin{cases}\ket{x}_S\ket{j}_M\ket{c}_C & \text{ if } c = 0\\
\ket{x'_j}_S\ket{j}_M\ket{c}_C & \text{ if } c = 1 \end{cases}\\
R&:\ket{j}_M\ket{c}_C \rightarrow \begin{cases} \ket{j}_M\ket{c}_C & \text{ if } (j,c) \neq (0,0) \\
-\ket{j}_M\ket{c}_C & \text{ if } (j,c) = (0,0)\end{cases}
}
\end{linenomath}
The operator $F$ is required to satisfy $F^2 = I$, which imposes constraints on the update moves.

The Coin register consists of a single qubit, while the number of qubits in the System and Move registers is problem-dependent. The quantum walk operator $W$ is then defined as
\begin{linenomath}
\eql{
W &= RV^\dagger B^\dagger  F B V 
}
\end{linenomath}
The quantum walk approach requires the ability to implement the controlled-$W$ ($\text{c}W$) operator, for which it suffices to be able to implement controlled versions of $R$ and $F$, since $\text{c}W = \lp \text{c}R\rp V^\dagger B^\dagger  \lp\text{c}F\rp B V$.

\subsubsection{Quantum Dihedral State Space}

\begin{figure}
    \centering
    \includegraphics[height=4.5cm]{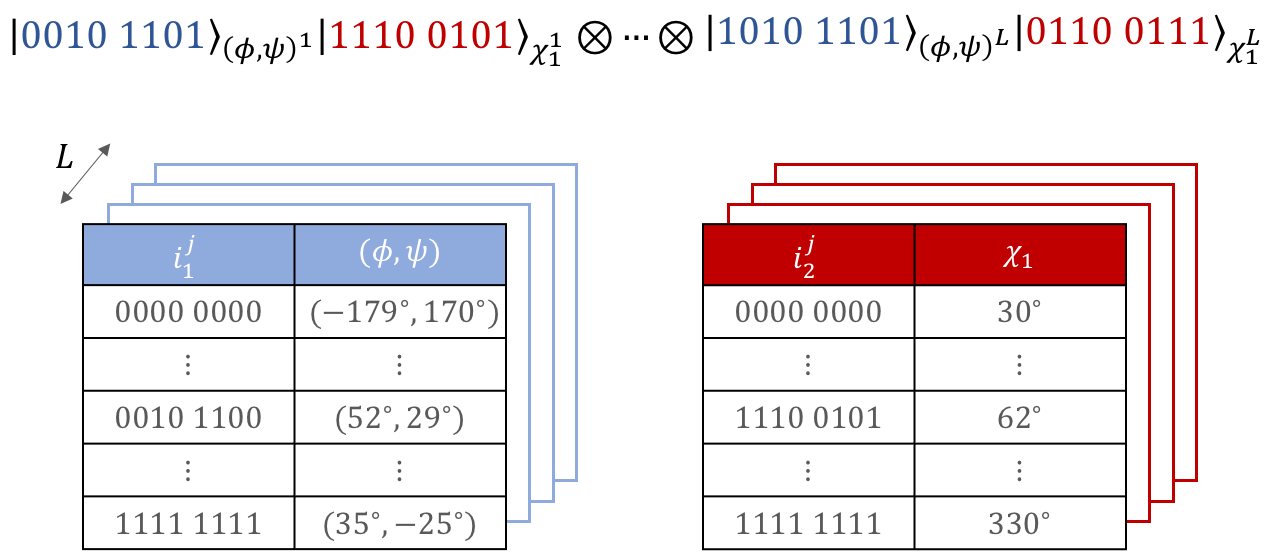}
    \caption{The state space of an $L$ residue loop consists of $2L$ registers. Each residue has a $(\varphi,\psi)$ register and a $\chi_1$ register, which store indices into lookup tables of dihedral angles sampled from appropriate Ramachandran and side chain libraries.}
    \label{fig:state-space}
\end{figure}

For simplicity of estimating resources we will assume that each side chain is described by a single dihedral angle $\chi_1$, although the generalization to side chains of varying lengths is straightforward. Assuming ideal bond lengths and angles, the state space is defined by $2L$ degrees of freedom given by the $(\varphi,\psi)^{j}$ backbone and $\chi^{j}_1$ side chain angles for each residue $j$\footnote{We treat $(\varphi,\psi)$ as a correlated pair and list the likely combinations of their values according to their joint distribution. Furthermore, due to the planar nature of the peptide bond, the $\omega$ angle is nearly always close to $180^\circ$, so as a first approximation we take $\omega$ to be fixed at this value.}.  For each residue, generate lookup tables $\mathfrak{T}^{j}_1$ for the $(\varphi,\psi)^{j}$ pairs and $\mathfrak{T}^{j}_2$ for the $\chi_1^{j}$ values by sampling data from backbone independent libraries. With these tables -- each of length $2^{b_T}$ -- populated, a given state of all of the backbone and side chain atoms in the H3 loop can be implicitly specified with $2Lb_T$ bits, by providing the indices to each of the lookup table values.  A quantum state on $2Lb_T$ qubits can then be specified as
\begin{linenomath}
\eql{
\bigotimes_{j=1}^L\ket{i_1^{j}}_{(\varphi,\psi)^j}\ket{i_2^{j}}_{\chi_1^j} \label{eq:state-of-all-residues}
}
\end{linenomath}
where $i_1^{j}$ and $i_2^j$ are the $b_T$-bit binary indices of $(\varphi, \psi)^j$ and $\chi^{j}_{1}$, respectively. This is a tensor product of $2L$ quantum registers, with odd registers corresponding to $(\varphi,\psi)$ and even registers corresponding to $\chi_1$ (Fig.\ref{fig:state-space}).

\subsubsection{Update Rule}
\label{sec:updrule}
We define the Markov chain update rule as: (i) Randomly select a residue $j$ from $1,\ldots, L$. (ii) Given the chosen residue, randomly select the backbone register or the side chain register. (iii) Add a uniformly random bit string to the register (i.e. bitwise XOR). The corresponding set $\+M$ of rules has size $\abs{\+M} = 2\cdot L \cdot 2^{b_T}$, and hence can be specified by a move register $\ket{j,k,\ell}_M = \ket{j}_{M_j}\ket{k}_{M_k}\ket{\ell}_{M_{\ell}}$ of $\log \abs{\+M} = b_T + \log L + 1$ qubits (ignoring rounding for notational convenience).

\textbf{$V$ operator.} Setting a uniform probability distribution over all possible moves, the $V$ operator then satisfies:
\begin{linenomath}
\eq{
V\ket{0}_M &= \frac{1}{\sqrt{\abs{\+M}}}\sum_{i=1}^{\abs{\+M}}\ket{i}_M \\
&= \frac{1}{\sqrt{L2^{b_T+1}}}\sum_{j=1}^L \sum_{k=0}^1 \sum_{\ell = 1}^{2^{b_T}}\ket{j,k,\ell}_M
}
\end{linenomath}
This operator can be implemented in a single step by applying Hadamard gates to all qubits, and does not require any Toffoli gates.

\textbf{c$F$ operator.} The $F$ operator acts on the state ($(\psi,\varphi),\chi_1$), move (M) and coin (C) registers as:
\begin{linenomath}
\eq{
F&\bigotimes_{r=1}^L\ket{i^r_1}_{(\psi,\varphi)^r}\ket{i^r_2}_{\chi_1^r} \ket{j,k,\ell}_M\ket{c}_C = \\
&\bigotimes_{r\neq j}\ket{i^r_1}_{(\psi,\varphi)^r}\ket{i^r_2}_{\chi_1^r} \ket{j,k,\ell}_{M}\ket{c}_C   \otimes \begin{cases}
 \ket{i^j_1 \oplus e_\ell\delta_{k,0}}_{(\psi,\varphi)^j}\ket{i^j_2 \oplus e_\ell\delta_{k,1}}_{\chi_1^j}& (c = 1) \\
 \ket{i^j_1}_{(\psi,\varphi)^j}\ket{i^j_2 }_{\chi_1^j} & (c = 0)
\end{cases}
}
\end{linenomath}
where $e_{\ell}$ is the $b$-bit binary representation of $\ell$, $\oplus$ is bitwise binary addition and $\delta_{i,j}$ is the Kronecker delta (applied bitwise). The $\text{c}F$ operator is a controlled indexed binary addition, with index given by the $\log L + 1$ qubits in the $k,\ell$ registers, i.e. if the control qubit is $1$, then the $b_T$ bits of the lookup table specified by $k,\ell$ are each XORed with the corresponding state register. By the unary iteration method of~\cite{babbush2018encoding} this can be implemented using $8L - 4 + 2Lb_T$ Toffoli gates and $\log L + 1$ temporary work qubits.

\textbf{c$R$ operator.} The $R$ operator acts as:
\begin{linenomath}
\eq{
R: \ket{j,k,\ell}_M\ket{c}_C &\rightarrow (-1)^{\delta_{jkl,000}\delta_{c,0}}\ket{j,k,\ell}_M\ket{c}_C
}
\end{linenomath}
where the $M$ and $C$ registers have a combined size of $\log L + b_T + 2$ qubits.  Adding an additional control qubit makes this equivalent (up to local Clifford operations) to a multiply-controlled $Z$ operation acting on the coin, with $\log L + b_T + 2$ controls. This can be implemented with $2\lp \log L + b_T+ 1\rp$ Toffoli gates~\cite{nielsen2002quantum}, using $\log L + b_T + 1$ temporary work qubits.

\textbf{B operator.} The $B$ operator performs a controlled rotation on the Coin register, conditioned on the state $\ket{x}$ of the system and the move $\ket{j}$ selected.
\begin{linenomath}
\eql{
\ket{x}_S\ket{j}_M \ket{0}_C &\ra \ket{x}_S\ket{j}_M \lp \sqrt{1-A_{x'_jx}}\ket{0}_C + \sqrt{A_{x'_jx}}\ket{1}_C\rp \label{eq:coherent-MH}
}
\end{linenomath}
In our case the state of the system is given by $\ket{x}_S = \bigotimes_{r=1}^L\ket{i_1^{r}}_{(\psi,\varphi)^r}\ket{i_2^r}_{\chi_1^r} $, and a proposed move is specified by three indices $\ket{j,k,\ell}_M$. $A_{x'x}$ is defined with respect to a particular energy function (see eq.~\eqref{eq:MH} of the main text), for which we use a classical all-atom force field.

\begin{figure}
    \centering
    \includegraphics[height=7cm]{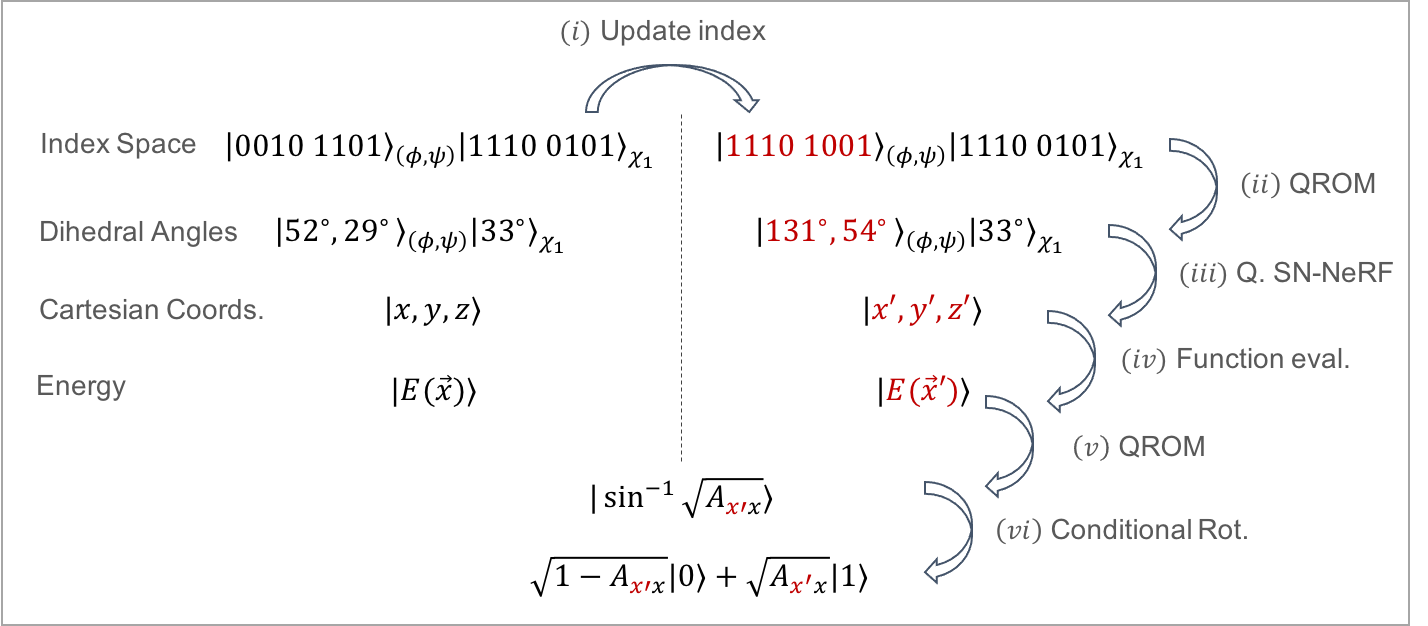}
    \caption{Steps required to implement the $B$ operator.  Left/right sides correspond to configuration before/after proposed update, with red indicating the changes due to the update.  Note that in the first three rows of this diagram, the quantum state of only a single residue is displayed. In reality the state of the system is a tensor product of registers for all $L$ residues (c.f. Eq.\eqref{eq:state-of-all-residues} and Fig.\ref{fig:state-space}). Following the conditional rotation, temporary registers are uncomputed.}
    \label{fig:B-operator}
\end{figure}

The transformation Eq.\eqref{eq:coherent-MH} can be decomposed into a number of steps (see Fig.\ref{fig:B-operator}):
\begin{linenomath}
\eq{
&  \ket{\psi_0}=\bigotimes_{r=1}^L \ket{i_1^{r}}_{(\varphi,\psi)^r}\ket{i_2^r}_{\chi_1^r} \ket{j,k,\ell}_M \\
\xrightarrow{(i)} & \ket{\psi_1} \defeq \ket{\psi_0} \ket{i^j_1 \oplus e_\ell \delta_{k,0}}_{(\varphi,\psi)^j}\ket{i^j_2\oplus e_\ell\delta_{k,1}}_{\chi_1^j}\\
\xrightarrow{(ii)} & \ket{\psi_2}\defeq   \ket{\psi_1}\bigotimes_{r=1}^L\ket{(\varphi,\psi)[i^r_1]\,}\ket{\chi_1[i^r_2]\,}\ket{(\varphi,\psi)[i^j_1 \oplus e_\ell\delta_{k,0}]}\ket{\chi_1[i_2^j \oplus e_\ell\delta_{k,1}]} \\
\xrightarrow{(iii)} & \ket{\psi_3}\defeq \ket{\psi_2}\bigotimes_{n=1}^N \ket{\vec{v}_n}\ket{\vec{v}'_n}\\
\xrightarrow{(iv)} & \ket{\psi_4}\defeq \ket{\psi_3}\ket{E(x)}\ket{E(x')}\\
\xrightarrow{(v)} & \ket{\psi_5}\defeq \ket{\psi_4}\ket{\theta_{x'x}}\\
\xrightarrow{(vi)} & \ket{\psi_6}\defeq \ket{\psi_5}\lp\sqrt{1-A_{x'_jx}}\ket{0} + \sqrt{A_{x'_jx}}\ket{1}\rp \\
\xrightarrow{(vii)} & \bigotimes_{r=1}^L\ket{i_1^{r}}_{(\varphi,\psi)^r}\ket{i_2^r}_{\chi_1^r} \ket{j,k,\ell}_M \lp\sqrt{1-A_{x'_jx}}\ket{0} + \sqrt{A_{x'_jx}}\ket{1}\rp
}
\end{linenomath}
In words: (i) Compute the updated $(\psi,\varphi)$ or $\chi_1$ indices in the register corresponding to residue $j$, as specified by the move. (ii) Look up the $(\psi,\varphi)$ and $\chi_1$ values corresponding to each index register from the appropriate lookup tables $\mathfrak{T}^j_1$ and $\mathfrak{T}^j_2$. (iii) Use the Quantum SN-NeRF algorithm (details below) to convert the dihedral angles for each residue to Cartesian coordinates for all the atoms in the loop, for both the original state $x$ of the chain and the updated state $x'_j$. (iv) Coherently evaluate $E(x)$ and $E(x'_j)$, which can be computed from the Cartesian coordinates of the atoms. (v) Use a function lookup table to evaluate $\theta_{x'x} = \arcsin\sqrt{A_{x'x}}=\arcsin\sqrt{\min\left\{1, e^{-(E(x') - E(x))/T}\right\}}$. (vi) Perform a conditional rotation on a single ancilla qubit based on $\theta_{x'x}$. (vii) Uncompute unneeded registers.

We next estimate the cost of implementing $B$ by the dominant contributions, which come from the conversion to 3D coordinates, and the non-bonding contributions to the energy function.

\subsubsection{Evaluating the force field coherently}

The CHARMM family of potential energy functions takes the form:
\eql{
E_{\text{CHARMM}} &= \sum_{bonds}k_b (b-b_0)^2 + \sum_{angles}k_{\theta}(\theta - \theta_0)^2 + \sum_{dihedrals} k_\varphi \left[ 1+ \cos(n\varphi-\delta)\right]  \nonumber \\
&+ \sum_{impropers}k_\omega(\omega-\omega_0)^2 + \sum_{Urey-Bradley}k_u(u-u_0)^2 \nonumber \\
&+ \sum_{non-bonded}  \lp \epsilon\left[\lp\frac{R_{ij}}{r_{ij}}\rp^{12} - \lp\frac{R_{ij}}{r_{ij}}\rp^{6}\right] +\frac{q_iq_j}{\epsilon r_{ij}}\rp \label{eq:ff}
}
where $b$ is the bond length between neighbouring bonded atoms, $\theta$ is the bond angle defined by three atoms, $\varphi$ is the dihedral angle defined by four atoms, $\omega$ is the \textit{improper} angle defined by four atoms (analagous to dihedral angles, but defined for four atoms arranged in a tetrahedron as opposed to in a linear chain), $u$ is the distance between atoms \ch{A} and \ch{C} in a bonded triple \ch{A}-\ch{B}-\ch{C} and $r_{ij}$ is the distance between non-bonded atoms $i$ and $j$.  The remaining symbols are parameters defined by the force field and are both experimentally and computationally determined. 

The evaluation of all-atom force fields such as \eqref{eq:ff} can be time-consuming due to the non-bonding terms

\begin{figure}
    \centering
    \includegraphics[height=8cm]{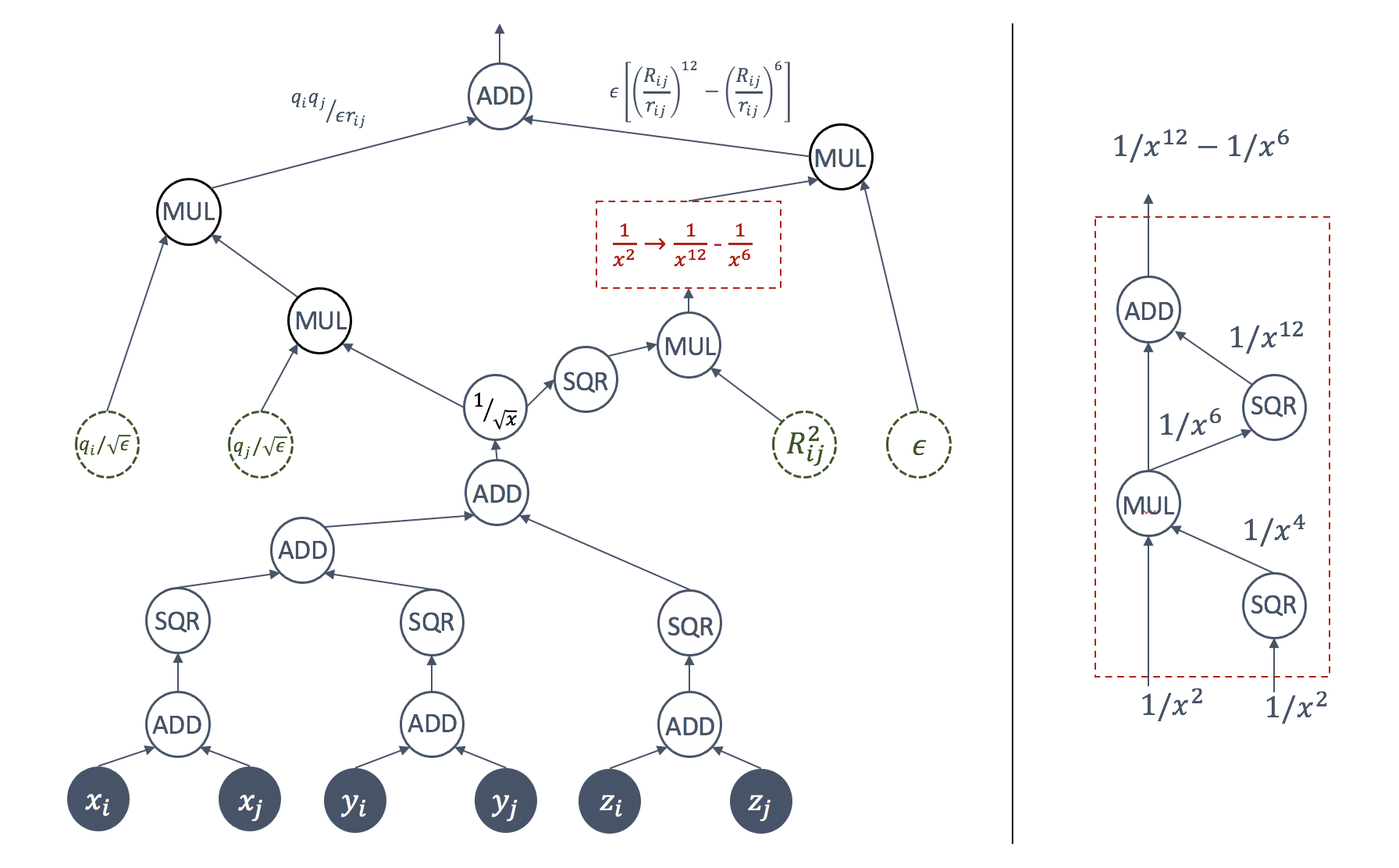}
    \caption{(Left) Computational graph for the single non-bonded term given by Eq.\eqref{eq:sum-non-bonded}. The red dashed box represents the sub-circuit which takes a value $1/x^2$ as input, and outputs $1/x^{12} - 1/x^6$. Cartesian coordinates of atoms $i$ and $j$ are input to $b$-bit accuracy (dark blue nodes).  Dashed nodes correspond to force-field parameters which can be directly compiled into the computational circuit gates, and do not need to be encoded in quantum registers.  (Right) Expansion of the dashed sub-circuit from the left side.}
    \label{fig:non-bond-comp-graph}
\end{figure}

\begin{linenomath}
\eql{
\sum_{i < j}\epsilon\left[\lp\frac{R_{ij}}{r_{ij}}\rp^{12} - \lp\frac{R_{ij}}{r_{ij}}\rp^{6}\right] +\frac{q_iq_j}{\epsilon r_{ij}} \label{eq:sum-non-bonded}
}
\end{linenomath}
where the sum is over all atoms $i,j$ in the loop.  As these terms dominate the time required to evaluate the overall force field, we estimate the quantum resources required by the resources needed for the non-bonding terms only. Fig~\ref{fig:non-bond-comp-graph} is a decomposition of a single term in this expression into a computational graph of elementary arithmetic expressions. The corresponding quantum circuit has Toffoli count of
\begin{linenomath}
\eq{
\+T = 7\+T_{\text{ADD}} + 5\+T_{\text{MUL}} + 6\+T_{\text{SQR}} + \+T_{\text{INVSQRT}} = 31b^2 + 7b
}
\end{linenomath}
using the values from Table~\ref{tab:toffoli-count}. The computational graph has $19$ nodes, and thus $19$ $b$-qubit ancillary registers are required to store the intermediate values and output. A loop with $N$ atoms therefore reuquires a total of $\frac{1}{2}N(N-1) (31b^2 + 8b)$ Toffoli gates and $\frac{20}{2}N(N-1)$ intermediate values (the additional factor of $b$ in the Toffoli gates and additional factor of $N(N-1)/2$ in the intermediate values is from the addition of all the pairwise terms together).

\textbf{Quantum SN-NeRF: Converting from dihedral to Cartesian coordinates.}\label{ss:q-SN-NeRF}
Here we give a quantum procedure for coherently converting from dihedral angles to atomic coordinates on a quantum computer, based on the classical SN-NeRF algorithm (see Algorithm~\ref{alg:NeRF}), itself a variant of the NeRF (Natural-extension Reference Frame) algorithm used by Rosetta.  Our quantum procedure applies the SN-NeRF steps coherently in quantum register, i.e. we take as input the state $\ket{\psi_{ABC}}:=\ket{\vec{a}}\ket{\vec{b}}\ket{\vec{c}}\ket{\cos\varphi}\ket{\sin\varphi}\ket{R_{cd}\cos\theta}\ket{R_{cd}\sin\theta}\ket{R_{bc}}$, append five zero-registers, each of length $3b$ bits ($b$-bits per $x,y,z$ coordinate) and perform:
\begin{linenomath}
\eq{
\ket{\psi_{ABC}}\ket{0^b}^{\otimes 5}&\rightarrow\ket{\psi_{ABC}}\ket{\vec{d'}}\ket{M_x} \ket{0^b}^{\otimes 3}\\
&\rightarrow \ket{\psi_{ABC}}\ket{\vec{{d'}}}\ket{M_x} \ket{M_z}\ket{0^b}^{\otimes 2} \\
&\rightarrow \ket{\psi_{ABC}}\ket{\vec{d'}}\ket{M_x} \ket{M_z} \ket{M_y} \ket{0^b}\\
&\rightarrow \ket{\psi_{ABC}}\ket{\vec{d'}}\ket{M_x} \ket{M_z} \ket{M_y} \ket{\vec{d}}\\
&\rightarrow \ket{\psi_{ABC}}\ket{\vec{d}}\ket{0^b}^{\otimes 4}
}
\end{linenomath}
where $M_{x,y,z}$, $\vec{d'}$ and $\vec{d}$ are defined as in Algorithm \ref{alg:NeRF}. The first four steps reproduce the classical SN-NeRF computations in quantum register, and the final step uncomputes unneeded registers.

\begin{Algorithm}
\hrulefill
\caption{Self-Normalizing Natural Extension Reference Frame (SN-NeRF~\cite{parsons2005practical})}
\hrulefill
\label{alg:NeRF}

\begin{algorithmic}
\STATE \textbf{Input: }Cartesian coordinates $\vec{a}, \vec{b}, \vec{c}$, dihedral angle $\varphi_{abcd} \equiv \varphi$, bond angle $\theta =\angle_{bcd}$, bond distances $R_{bc}, R_{cd}$.
\STATE
\STATE $\vec{d'} \gets \left[ R_{cd}\cos\theta, R_{cd}\cos\varphi\sin\theta, R_{cd}\sin\varphi\sin\theta\right]^\top$
\STATE $M_x\gets \hat{bc} := \frac{\vec{c} - \vec{b}}{R_{bc}}$
\STATE $M_z\gets \hat{n} := \frac{\lp \vec{b} - \vec{a}\rp \times \hat{bc}}{\abs{\lp \vec{b} - \vec{a}\rp \times \hat{bc}}}$
\STATE $M_y\gets \hat{n}\times \hat{bc}$
\STATE $\vec{d}\gets \left[ M_x, M_y, M_z\right]\cdot \vec{d'} + \vec{c}$
\STATE
\STATE  \textbf{Output: } Cartesian coordinates $\vec{d}$.
\end{algorithmic}
\hrulefill
\end{Algorithm}

The inputs to Algorithm~\ref{alg:NeRF} are vectors $\vec{a}, \vec{b},\vec{c}$ containing the 3D Cartesian coordinates of atoms $A,B,C$, the fixed angle $\theta$ between atoms $B,C,D$, and the dihedral angle $\varphi$ defined by atoms $A,B,C,D$.  We assume the inputs to $\ket{\psi_{ABC}}$ are readily available in quantum registers\footnote{We assume fixed ideal bond angles $\theta$, so $R_{cd}\cos\theta$ and $R_{cd}\sin\theta$ can be precomputed and stored in register. Instead of reading $\varphi$ from a lookup table and then computing the $\sin$ and $\cos$ values coherently, one can instead also precompute and store these values and store them in lookup tables directly.}.

\begin{figure}
    \centering
    \includegraphics[height=8cm]{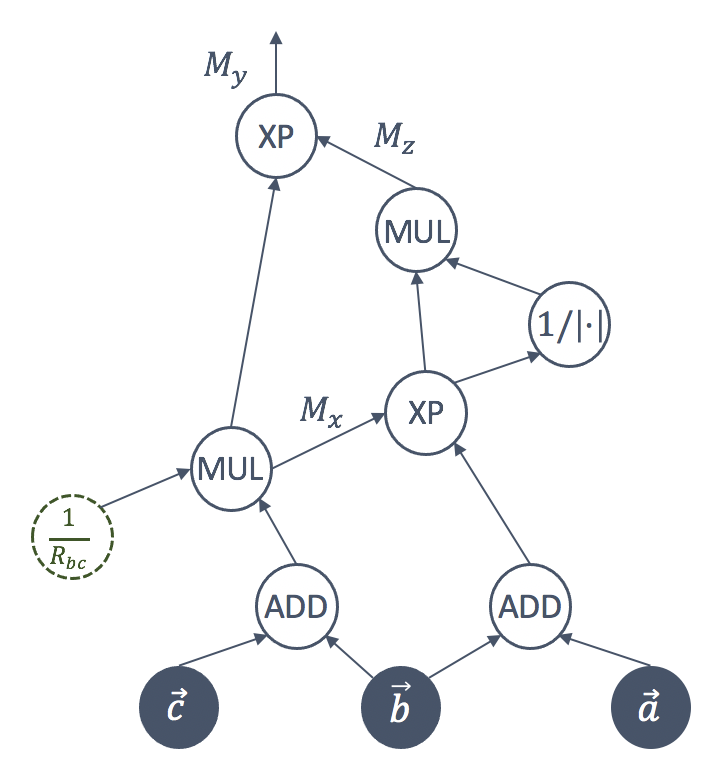}
    \caption{Quantum SN-NeRF: computational graph for computing $M_x, M_y, M_z$ coherently. Each of the nodes $\vec{a}$, $\vec{b}$, $\vec{c}$ represent three registers of $b$ qubits each.}
    \label{fig:QSN-NeRF-comp-graph}
\end{figure}

In the first step of the quantum algorithm, the state $\ket{d'}=\ket{d'_1}\ket{d'_2}\ket{d'_3}$ corresponding to the components of $d'$ can be prepared via the transformation
\begin{linenomath}
\eq{
\ket{R_{cd}\sin\theta}\ket{R_{cd}\cos\theta}\ket{\cos \varphi}\ket{\sin \varphi}&\ra \ket{R_{cd}\sin\theta}\ket{\cos\varphi}\ket{\sin\varphi}\ket{R_{cd}\cos\theta}\ket{R_{cd}\cos\varphi\sin\theta}\ket{R_{cd}\sin\varphi\sin\theta}\\
&= \ket{R_{cd}\sin\theta}\ket{\cos\varphi}\ket{\sin\varphi}\ket{d'_1}\ket{d'_2}\ket{d'_3}
}
\end{linenomath}
which can be completed with $\+T_d' = 2\+T_{\text{MUL}}$ Toffoli gates. Then, noting that for two $3$-dimensional vectors $\vec{u}, \vec{v}$ stored in quantum register, the reciprocal norm $1/\sqrt{\sum_i u_i^2}$ and the cross product $\vec{u}\times \vec{v}$ can be computed with Toffoli counts $T_{1/\norm{\cdot}}=3\+T_{\text{SQR}} + 2\+T_{\text{ADD}} + \+T_{\text{INVSQRT}}$ and $T_\text{XP}=6\+T_{\text{MUL}} + 3\+T_{\text{ADD}}$, respectively, $M_x, M_y, M_z$ can be implemented with Toffoli counts
\begin{linenomath}
\eq{
M_x: &\quad 3 \+T_\text{ADD} + 3\+T_{MUL} \\
M_y: &\quad \+T_\text{XP} = 6\+T_\text{MUL} + 3\+T_\text{ADD}\\
M_z: &\quad 3\+T_\text{ADD}+\+T_\text{XP} + \+T_{1/\norm{\cdot}} + \+T_\text{MUL}= 7\+T_\text{MUL} + 3\+T_\text{SQR} + 8\+T_\text{ADD} + \+T_\text{INVSQRT}
}
\end{linenomath}
via quantum circuit corresponding to the computational graph in Fig.\ref{fig:QSN-NeRF-comp-graph}. With $\ket{d'}$ in register, the output vector $d = Md' + c$ can then be computed with an additional $9\+T_\text{MUL} + 3\+T_\text{ADD}$ Toffolis.  Counting operations, the total Toffoli count needed to evaluate $d$ in quantum register is 
\begin{linenomath}
\eq{
\+T &= 27\+T_{\text{MUL}} + 17\+T_{\text{ADD}} + 3\+T_{\text{SQR}} + T_{\text{INVSQRT}} 
= 73.5b^2 + 17 b 
}
\end{linenomath}
for $b$-bit operations, using values from Table \ref{tab:toffoli-count}, and requires $48$ ancilla registers of $b$ qubits each. To compute the coordinates of the $N$ heavy atoms in the loop thus requires approximately $N\+T$ Toffoli gates and $48N$ registers, each of $b$ qubits.

\section{Benchmarking}\label{app:benchmarking}

We benchmark our approach against the $L=14$ residue H1 loop MKHMAGAAAAGAVV from the Syrian hamster prion protein, which was modelled in~\cite{ulmschneider2006monte} using $4\times 10^{9}$ classical MC steps over $10.6$ days. A structural snap-shot was taken every $N_c =10^6$ MC steps, for a total of $4\times 10^3$ snapshots. This protein has $N=88$ atoms (excluding hydrogens). Using these values and the information in Table~\ref{tab:resources-summary} of the main text, the controlled quantum walk operator 
$\text{c}W = (\text{c}R)V^\dagger B^\dagger (\text{c}F) BV$ 
can be seen to require $1.3\times 10^8$ Toffoli gates and $3.2\times 10^5$ $b$-qubit registers to store temporary arithmetic values (corresponding to the number of elementary arithmetic operations involved). Table~\ref{tab:adjusted-toffoli} gives an estimate of the adjusted Toffoli count and number of temporary ancilla registers required based on the uncomputation depth $\ell$, where we approximate the scaling of the number of Toffoli gates by the scaling of the total number of arithmetic operations with $\ell$.    The final resource estimates for modelling the hamster prion H1 loop on a quantum computer are given in Table~\ref{tab:benchmark-summary}, where we assume that the spectral gap of the classical Markov chain transition matrix can be estimated by $\Delta = \frac{1}{N_c}$, and thus the number of quantum MC steps required per structural sample is $N_q = \sqrt{\frac{2}{\Delta}}= \sqrt{2 N_c} = 1.4\times 10^3$.  

\begin{table}
\centering
\resizebox{\textwidth}{!}{%
\begin{tabular}{l|c|c}
$\+T_{W}$ (Toffoli per $W$) & $62N^2 b^2 + 232Nb^2 + 52Nb + 16N^2b + O(Lb_T)$ & $1.3\times 10^8$\\ \hline
$N_{ops}$ (arithmetic operations per $W$) & $40N^2 + 152N + O(\log L)$ & $3.2\times 10^5$\\ \hline
$c_u$ (uncomputation overhead)~\cite{bennett1989time} &  $N_{ops}^{\log(2\ell-1)/\log \ell -1}$& $\star$\\ \hline 
adjusted Toffoli count per $W$ & $c_u\+T_{W}$ & $\star$\\ \hline
$b$-qubit ancilla registers required~\cite{bennett1989time} & $\frac{\log N_{ops}}{\log \ell}(\ell-1)+1$ & $\star$
\end{tabular}%
}
\caption{Number of ancilla registers and adjusted number of Toffoli gates per $W$ operator, based on the number of arithmetic operations $\ell$ carried out before each uncomputation. Right hand column corresponds to the values $N=88$, $L=14$, $b=16$, $b_T=8$. $(\star)$ values are $\ell$-dependent (see Table~\ref{tab:benchmark-summary}).}
\label{tab:adjusted-toffoli}
\end{table}

\begin{table}
\resizebox{\textwidth}{!}{%
\begin{tabular}{l|c|c|c|c}
& Gen.1 TF-1 & Gen.1 TF-100 & Future TF-100 & Future TF-1000 \\ \hline
$\ell$ (uncomputation depth) & $75$ & $155$ & $8,000$ & $100,000$ \\
$c_u$ (uncomputation overhead)  & $7.5$ & $5.7$ & $2.7$ & $2.2$ \\ 
unadjusted Toffoli count per $W$ & $1.3\times 10^8$& $1.3\times 10^8$& $1.3\times 10^8$& $1.3\times 10^8$\\  \hline
adjusted Toffoli count per $W$ & $9.8\times 10^8$ & $7.4\times 10^8$ & $3.5\times 10^8$ & $2.8\times 10^8$\\
$N_q$ (quantum applications of $W$) & $1.4\times 10^3$ & $1.4\times 10^3$ & $1.4\times 10^3$ & $1.4\times 10^3$\\
total Toffoli count per factory ($N_q$ iterations) & $1.4\times 10^{12}$ & $1.0\times 10^{10}$ & $4.9\times 10^{9}$ & $4.0\times 10^{8}$\\
Toffoli distillation rate / $\mu$s~\cite{gidney2019efficient} & $170.5$& $170.5$ & $12.1$ & $12.1$\\ 
total distillation time / structural sample& $7.5$ years & $20.6$ days & $16.5$ hours & $1.3$ hours\\ \hline
logical $b$-bit ancilla registers & $3.5\times 10^3$ & $6.2\times 10^3$ & $1.8\times 10^5$ & $1.8\times 10^6$ \\
physical qubits / logical data qubit & $1,083$ & $1,083$ & $507$  & $507$\\
physical ancilla qubits required & $3.8\times 10^6$ & $6.7\times 10^6$ & $9.1\times 10^7$ & $8.9\times 10^8$\\ \hline
physical qubits / factory & $132,528$ & $132,528$ & $19,632$ & $19,632$\\
total physical factory qubits & $1.3\times 10^5$ & $1.3\times 10^7$ & $1.9\times 10^6$ & $1.9\times 10^7$ \\ \hline
total physical qubits & $3.9\times 10^6$ & $2.0\times 10^7$ & $9.4 \times 10^7$ & $9.1\times 10^8$
\end{tabular}%
}
\caption{Estimated computational time and physical qubits required for modelling the H1 prion protein of~\cite{ulmschneider2006monte} on first generation large-scale FTQC as well as future devices. TF-X indicates that X Toffoli factories are available for use in parallel. Values of uncomputation depth $\ell$ are chosen to balance computational time with number of qubits required. Table computed with the aid of the resource estimation spreadsheet from the supplementary information of~\cite{gidney2019efficient}.}
\label{tab:benchmark-summary}
\end{table}

\end{document}